\documentclass[fleqn]{article}
 
 \def\const{{\rm constant}}
 \date{}
 
\begin{document}
 
 \title{Exact radiative spacetimes: some recent developments}
 \author{Ji\v r\' \i\ Bi\v c\' ak}
 \newcommand{\keywords}{Radiative solutions, Einstein's field equations} 
 %%%%%%%%%%%%%%%%%%%%%%%%%%%%%%%%%%%%%%%%%%%%%%%%%%%%%%%%%%%%%%%%%%%%%%%%%%%%%%
 %%%%%%%%%%%%%%%% please give up to three PACS numbers here %%%%%%%%%%%%%%%%%%%
 %%%%%%%%%%%%%%%%%%%%%%%%%%%%%%%%%%%%%%%%%%%%%%%%%%%%%%%%%%%%%%%%%%%%%%%%%%%%%%
 \newcommand{\PACS}{04.20.Jb, 04.30.-w, 04.70.Bw}
 %%%%%%%%%%%%%%%%%%%%%%%%%%%%%%%%%%%%%%%%%%%%%%%%%%%%%%%%%%%%%%%%%%%%%%%%%%%%%%
 %% please enter (First) Author (et al.) and short version of the title here %%
 %%%%%%%%%%%% must not exceed 80 characters in length together %%%%%%%%%%%%%%%%
 %%%%%%%%%%%%%%%%%%%%%%%%%%%%%%%%%%%%%%%%%%%%%%%%%%%%%%%%%%%%%%%%%%%%%%%%%%%%%%
 \newcommand{\shorttitle}
 {Ji\v r\' \i\ Bi\v c\' ak, Exact radiative spacetimes} 
 %% sets the header on oddpage
 %
 %\authorrunning{Ji\v r\' \i\ Bi\v c\' ak}
 % if there are more than two authors,
 % please abbreviate author list for running head
 %
 %
 \newcommand{\address}
 {Institute of Theoretical Physics, Charles University,
 V Hole\v sovi\v ck\' ach 2, 180 00 Prague 8,
 Czech Republic}
 \newcommand{\email}{\tt bicak@mbox.troja.mff.cuni.cz} 
 \maketitle              % typesets the title of the contribution
 
 \begin{abstract}
 Five classes of radiative solutions of Einstein's field equations
 are discussed in the light of some new developments. These are
 plane waves and their collisions, cylindrical waves,
 Robinson-Trautman and type N spacetimes, boost-rotation symmetric spacetimes
 and generalized Gowdy-type cosmological models.
 \end{abstract}

 \vskip 2cm 
 
 \section{Introduction}
 Gernot Franz Sebastian Neugebauer suggested that I should give a
 fairly broad review on radiative spacetimes. Johann Wolfgang von
 Goethe believes that this is not easy: ``It is extremely difficult
 to report on the opinions of others ... If the reporter goes
 into detail, he creates impatience and boredom; if he wants to
 summarize, he risks giving his own point of view; if he avoids
 judgements, the reader does not know where to begin, and if he
 organizes his materials according to principles, the presentation
 becomes one-sided and arouses opposition, and the history itself
 creates new histories." (J.W.Goethe, ``Materialien zur Geschichte
 der Farbenlehre".)
 
 My ``Materialien to the Exact Gravitational-Waves Lehre" will be:
 plane waves and their collisions; cylindrical waves and null
 infinity in (2+1)-dimensional spacetimes; Robinson-Trautman
 solutions and type N twisting spacetimes; boost-rotation
 symmetric spacetimes and spinning C-metric; ``cosmological" waves
 (Gowdy models) and the approach to cosmological singularity.
 Some parts of the ``Materialien'' are taken from \cite{BISP}; see also
 \cite{KVB} for the more detailed review and references until 1995.

 \section{Plane waves and their collisions}
 
 By the definition (see e.g. \cite{KSH}) a
 vacuum spacetime is a {\it ``plane-fronted gravitational wave''} if it
 contains a shearfree geodesic null
 congruence (with tangents $k^\alpha$), and if it admits ``plane wave surfaces'' (spacelike
 2-surfaces orthogonal to $k^\alpha$). Because of the existence of plane
 wave surfaces, the expansion and twist  must vanish as well.
 The best known subclass of these waves are
 {\it ``plane-fronted gravitational waves with parallel rays''
 (pp-waves)} which are defined by the condition that the null
 vector $k^\alpha$ is covariantly constant, $k_{\alpha ; \beta} =
 0$. 
 
 In suitable null coordinates with
 a null coordinate $u$ such that $k_\alpha = u_{,\alpha}$ and
 $k^\alpha = \left(\partial/\partial v\right)^\alpha$, the metric has the form
 \begin{equation}
 \label{Equ51}
 ds^2 = 2d \zeta d \bar \zeta - 2 du dv - 2H(u, \zeta, \bar
 \zeta) du^2,
 \end{equation}
 where $H$ is a real function dependent on $u$, and on the
 complex coordinate $\zeta$ which spans the wave 2-surfaces
 $u = \const$, $v = \const$. The vacuum field
 equations imply 
 $2H = f(u, \zeta) + \bar f (u, \bar \zeta)$,
 where $f$ %(u, \zeta)$
 is an arbitrary function of $u$, analytic in $\zeta$. 
 
 In general the pp-waves have only the single isometry
 generated by the Killing vector $k^\alpha = \left(\partial/\partial v\right)^\alpha$.
 However, a much larger {group of symmetries} may exist for
 various particular choices of the function $H(u, \zeta, \bar \zeta)$.
 Jordan, Ehlers and Kundt \cite{JEK} (see also 
 \cite{KSH}) gave a complete classification of the pp-waves in
 terms of their symmetries and corresponding special forms of $H$.
 For example, in the best known case of plane waves 
 $H(u, \zeta, \bar \zeta) = A (u) \zeta^2 + \bar A (u) \bar \zeta^2$,
 with $A(u)$ being an arbitrary function of $u$. 
 This spacetime admits five Killing vectors.
 
 Recently, Aichelburg and Balasin \cite{AiB,AiB2}
 generalized the classification given in \cite{JEK} by admitting
 distribution-valued profile functions and allowing for
 non-vacuum spacetimes.
 They have shown that with $H$ in the form of
 delta-like pulses,
 $H(u, \zeta, \bar \zeta) = f(\zeta, \bar \zeta) \delta(u)$,
 new symmetry classes arise even in the vacuum case.
 
 The main motivation to consider impulsive pp-waves stems from
 the metrics describing a black hole or a ``particle'' boosted to
 the speed of light. The simplest metric of this type, given by
 Aichelburg and Sexl \cite{AS}, is a Schwarzschild black
 hole with mass $m$ boosted in such a way that $\mu = m/\sqrt {1-w^2}$ is
 held constant as $w \rightarrow 1$. It reads
 \begin{equation}
 ds^2 = 2d \zeta d \bar \zeta - 2 du dv - 4 \mu \log (\zeta \bar
 \zeta) \delta (u) du^2,
 \end{equation}
 with $H$ clearly in the form of a delta pulse. This is not a vacuum metric:
 the energy-momentum tensor $T_{\alpha \beta} = \mu \delta(u)
 \delta (\zeta)k_\alpha k_\beta$ indicates that there is a
 ``point-like particle'' moving with the speed of light along $u=0$.
 
 The interest in impulsive waves generated by boosting a
 ``particle'' at rest to the velocity of light by means of an
 appropriate limiting procedure persists up to the present. The
 ultrarelativistic limits of Kerr and Kerr-Newman black holes were
 obtained \cite{LoSa,FePe,BaNa}, and recently, boosted
 static multipole (Weyl) particles were studied \cite{PoGr1}.
 Impulsive gravitational waves were also generated by boosting 
 the Schwarzschild-de Sitter and  Schwarzschild-anti de Sitter
 metrics to the ultrarelativistic limit \cite{HoTa,PoGr2};
 see the contribution by J. Griffiths and J. Podolsk\'y in these Proceedings.
 
 These types of spacetimes, especially the simple Aichelburg-Sexl
 metrics, have been employed in current problems of the
 generation of gravitational radiation from 
 axisymmetric black hole collisions and black hole encounters. The recent
 monograph by d'Eath \cite{PE} gives a comprehensive survey,
 including the author's new results. There is good reason to believe
 that spacetime metrics produced in high speed collisions will be
 simpler than those corresponding to (more realistic) situations
 in which black holes start to collide with low relative
 velocities. The spacetimes corresponding to the collisions at
 exactly the speed of light is an interesting limit which can be
 treated most easily. Aichelburg-Sexl metrics are used to describe
 limiting ``incoming states'' of two black holes, moving one
 against the other with the speed of light. 
 Great interest has been stimulated by 't Hooft's \cite{Hoo} work
 on the quantum scattering of two pointlike particles at
 centre-of-mass energies higher or equal to the Planck energy.
 This quantum process has been shown to have close connection with
 classical black hole collisions at the speed of light 
 (see \cite{PE,Fab} and references therein).
 
 Recently, the Colombeau algebra of generalized functions, which
 enables one to deal with singular products of distributions, has
 been brought to general relativity and used in the description
 of impulsive pp-waves in various coordinate systems
 \cite{Kus1}, and also for a rigorous solution of the geodesic and
 geodesic deviation equations for impulsive waves \cite{KuS}.
 The investigation of the equations of geodesics in
 non-homogeneous pp-waves (with $f \sim \zeta ^3$) has shown
 that the motion of test particles is chaotic 
 (see \cite{PoVe} and the contribution by J. Podolsk\'y in these Proceedings).
 
 Plane-fronted waves have been used as simple metrics in various
 other contexts, for example, in quantum field theory on a given
 background (see e.g. \cite{Per}). As emphasized very recently by Gibbons
 \cite{Gi}, since for pp-waves and type N Kundt's class 
 all possible invariants formed from the
 Weyl tensor and its covariant derivatives vanish \cite{BiPr},
 these metrics suffer no quantum corrections to all loop orders.
 Thus they may offer insights into the behaviour of a full
 quantum theory. 
 
 \subsection*{Colliding plane waves}
 
 The first detailed study of colliding plane waves was undertaken independently
 by Khan and Penrose and by Szekeres (see
 \cite{KSH,Gr} for references). Szekeres formulated the problem as a
 characteristic initial value problem for a system of hyperbolic
 equations in two variables (null coordinates) $u, v$ with data
 specified on the pair of null hypersurfaces, say $u=0, v=0$
 intersecting in a spacelike 2-surface. Although Szekeres' formulation 
 of a general solution for the
 problem of colliding parallel-polarized waves is difficult to use
 for constructing explicit solutions, it has been
 employed in a general analysis of the structure of the
 singularities produced by the collision \cite{Yu1}.
 
 It has also inspired the work developed at the beginning of
 the 1990s by Hauser and Ernst \cite{HET}.
 Their new method of analyzing the initial value problem can be
 used also when the polarization of the approaching
 waves is not aligned. They formulated the initial value problem in terms
 of the equivalent matrix Riemann-Hilbert
 problem. Their techniques are related to
 those used by Neugebauer and Meinel to construct and analyze the
 rotating disk solution as a boundary value problem 
 (see their contributions to these Proceedings).
 Most recently, Hauser and Ernst prepared an 
 extensive treatise \cite{HET1} in which they give a general description
 and detailed mathematical proofs of their study of the solutions of the hyperbolic 
 Ernst equation.
 
 The papers on colliding plane waves published until 1991 are reviewed in
 \cite{Gr} (see also \cite{KVB}). New developments 
 have been mostly involved with ``non-classical'' issues like the inclusion of
 dilatonic fields or the discussion of the particle production.
 One of the few exceptions has been the analysis of colliding waves in the expanding
 backgrounds as e.g. in Friedmann-Robertson-Walker universes filled  by stiff fluid \cite{BIJG}.
 In contrast to the waves propagating and colliding on the ``flat backgrounds'',
 no singularities arise in the expanding backgrounds.

 %%%%%%%%%%%%%%%%%%%%%%%%%%%%%%%%%%%%%%%%%%%%%%%%%%%%%%%%%%%%%%%%%%%%%%%%
 \section{Cylindrical waves}
 %%%%%%%%%%%%%%%%%%%%%%%%%%%%%%%%%%%%%%%%%%%%%%%%%%%%%%%%%%%%%%%%%%%%%%%%
 
 Despite the fact that cylindrically symmetric waves cannot
 describe exactly the radiation from bounded sources, they even recently played
 an important role in clarifying a number of complicated issues, 
 such as testing the quasilocal mass-energy \cite{To}, testing
 codes in numerical relativity \cite{RI}, investigation of the
 cosmic censorship  \cite{BER}, and quantum gravity \cite{AP} 
 (see \cite{BISP} for more details and references).
 
 In recent work with Ashtekar and Schmidt \cite{ABS1,ABS2},
 we considered gravitational waves with a
 space-translation Killing field (``generalized Einstein-Rosen
 waves''). In the (2+1)-dimensional framework the
 Einstein-Rosen subclass forms a simple instructive example of
 explicitly given spacetimes which
 admit a smooth global null (and timelike) infinity even for
 strong initial data. 
 
 4-dimensional vacuum gravity which admits a spacelike hypersurface Killing vector 
 $\partial/\partial{z}$ is equivalent to 3-dimensional
 gravity coupled to a scalar field. In 3 dimensions, there is no
 gravitational radiation. Hence, the local degrees of freedom are
 all contained in the scalar field. One therefore expects that 
 Cauchy data for the scalar field will suffice to determine the
 solution. For data which fall off appropriately, we thus expect
 the 3-dimensional Lorentzian geometry to be asymptotically flat
 in the sense of Penrose, i.e. that there
 should exist a 2-dimensional boundary representing null infinity.
 In general cases, this is analyzed in \cite{ABS1}. 
 
 Restricting here ourselves to the Einstein-Rosen waves by assuming that
 there is a further spacelike, hypersurface orthogonal Killing
 vector $\partial/\partial \varphi$ which commutes with
 $\partial/\partial z$, we find the 3-metric given by
 \begin{equation}
 \label{Equ70}
 d\sigma^2= g_{ab}dx^adx^b = e^{2\gamma}(-dt^2+d\rho^2)+
 \rho^2 d\varphi^2 ,
 \end{equation}
 where $\gamma = \gamma(t,\rho)$.
 The field equations for the scalar field $\psi$ coupled to 
 this metric become
 \begin{equation}
 - \ddot\psi + \psi'' + \rho^{-1}\psi' = 0~,~~~
 \gamma' = \rho (\dot{\psi}^2+\psi'^2)~, ~~~
 \dot \gamma = 2\rho \dot \psi \psi'.
 \end{equation}
 Thus, we can first solve the axisymmetric wave equation
 for $\psi$ on Minkowski space and then solve for
 $\gamma$ -- the only unknown metric coefficient -- by quadratures.
 The ``method of descent'' from the Kirchhoff formula in 4
 dimensions
 gives the representation of the solution of the wave
 equation in 3 dimensions in terms of Cauchy data $\Psi_0 =
 \psi(t=0,x,y), \Psi_1 = \psi_{,t} (t=0,x,y)$ (see \cite{ABS1}).
 We assume that the Cauchy data are axially symmetric 
 and of compact support.
 
 Investigating the behaviour of the solution at future null
 infinity $\cal J^+$, one finds 
 \begin{equation}
 \label{Equ73}
 \psi (u, \rho) = \frac {f_0(u)}{\sqrt{\rho}} + \frac
 {1}{\sqrt{\rho}} \sum_{k=1}^{\infty} \frac{f_k(u)}{\rho^k}  ,
 \end{equation}
 where $u = t- \rho$ and the coefficients $f$'s are determined by the
 Cauchy data.
 The field equations imply
 \begin{equation}
 \label{Equ80}
 \gamma = \gamma_0 - 2 \int_{-\infty} ^{u}\left[\dot f_0(u)\right]^2
 du - \sum_{k=1}^{\infty} \frac{h_k(u)}{(k+1)\rho^{k+1}} .
 \end{equation}
 Thus, $\gamma$ also admits an expansion in $\rho^{-1}$.
 It is straightforward to show that the spacetime admits a smooth future null 
 infinity by setting $\tilde \rho= \rho ^{-1}, \tilde u = u, \tilde
 \varphi= \varphi$ and rescaling $g_{ab}$ by a conformal factor $\Omega
 = \tilde \rho$. 
 Hence, the (2+1)-dimensional curved spacetime
 has a smooth (2-dimensional) null infinity.  Penrose's picture
 works for arbitrarily strong initial data $\Psi_0$, $\Psi_1$.
 We can thus conclude that {\it cylindrical waves in
 (2+1)-dimensions give an explicit model of the 
 Bondi-Penrose radiation theory which admits smooth null and timelike 
 infinity for arbitrarily strong initial data}.
 There is no other such model available.
 The general results on the existence of $\cal J$ in 4 
 dimensions assume weak data.
 
 %%%%%%%%%%%%%%%%%%%%%%%%%%%%%%%%%%%%%%%%%%%%%%%%%%%%%%%%%%%%%%%%%%%%%%%%
 \section{On the Robinson-Trautman and type N twisting solutions}
 %%%%%%%%%%%%%%%%%%%%%%%%%%%%%%%%%%%%%%%%%%%%%%%%%%%%%%%%%%%%%%%%%%%%%%%%
 
 These spacetimes have attracted increased attention in the
 last decade -- most notably in the work by Chru\'{s}ciel, and
 Chru\'{s}ciel and Singleton \cite{ak}. In these studies the
 Robinson-Trautman spacetimes have been shown to exist globally for all
 positive ``times'', and to converge asymptotically to a
 Schwarzschild metric. Interestingly, the extension of the
 spacetimes across the ``Schwarz\-schild-like'' event horizon can only
 be made with a finite degree of smoothness. These 
 studies are based on the derivation and analysis of an asymptotic
 expansion describing the long-time behaviour of the solutions of
 the nonlinear parabolic Robinson-Trautman equation.
 
 In our recent work \cite{BiPo,al}, we studied Robinson-Trautman 
 spacetimes with a positive cosmological constant $\Lambda$. The
 results proving the global existence and convergence of the
 solutions of the Robinson-Trautman equation can be taken over
 from the previous studies since $\Lambda$ does not explicitly
 enter this equation. We have shown that,
 starting with arbitrary, smooth initial data at $u=u_0$,
 these cosmological Robinson-Trautman solutions converge exponentially fast
 to a Schwarzschild-de Sitter solution at large retarded times
 ($u\to \infty$).
 The interior of a
 Schwarzschild-de Sitter black hole can be joined to an ``external''
 cosmological Robinson-Trautman spacetime across the horizon
 $\cal H^+$ with
 a higher degree of smoothness than in the corresponding case with
 $\Lambda = 0$. In particular, in the extreme case with
 $9 \Lambda m^2 = 1$, in which the black hole and cosmological horizons coincide,
 the Robinson-Trautman spacetimes can be extended smoothly
 through $\cal H^+$ to the extreme Schwarzschild-de Sitter spacetime
 with the same values of $\Lambda$ and $m$. However, such an extension 
 is not analytic (and not unique). 
 
 We have also demonstrated that the cosmological
 Robinson-Trautman solutions are explicit models exhibiting
 the cosmic no-hair conjecture. As far as we are aware,
 these models represent the only exact analytic demonstration of the cosmic
 no-hair conjecture under the presence of gravitational waves.
 They also appear to be the only exact examples of a black hole
 formation in nonspherical spacetimes which are not asymptotically
 flat. 
 
 \subsection*{Type N twisting spacetimes}
 
 Since diverging, non-twisting Robinson-Trautman
 spacetimes of type N have singularities, there has been hope
 that if one admits a nonvanishing twist a more realistic
 radiative spacetime may exist.
 
 Stephani \cite{STP}, however, indicated, by constructing a general
 solution of the linearized equations, that singularities at infinity
 probably occur. More recently, Finley et al \cite{FIL} found
 an approximate twisting type N solution up to the third
 order of iteration on the basis of which they suggested that it
 seems that the twisting, type N fields can describe a
 radiation field outside bounded sources. However, employing
 the Newman-Penrose formalism and MAPLE we succeeded in discovering a
 nonvanishing quartic invariant in the 2nd derivatives of the
 Riemann tensor \cite{BiPr},
 which shows that solutions of both Stephani and Finley et al
 contain singularities at large $r$. Very recently, Mac Alevey
 \cite{MA} argued that an approximate solution at any finite order
 can be calculated without occurrence of singularities. It is very
 likely, however, that a corresponding exact solution must contain
 singularities since Mason \cite{MASO} proved that the
 only vacuum algebraically special spacetime that is
 asymptotically simple is the Minkowski space.
 
 Even if a radiative solution with a complete smooth null infinity
 may be out of reach, it is of interest to construct
 radiative solutions which admit at least a global null infinity in
 the sense that its smooth cross sections exist although this null
 infinity is not necessarily complete. The only explicit examples
 of such solutions are spacetimes with boost-rotation symmetry.

 %%%%%%%%%%%%%%%%%%%%%%%%%%%%%%%%%%%%%%%%%%%%%%%%%%%%%%%%%%%%%%%%%%%%%%%%
 \section{The boost-rotation symmetric radiative spacetimes}
 %%%%%%%%%%%%%%%%%%%%%%%%%%%%%%%%%%%%%%%%%%%%%%%%%%%%%%%%%%%%%%%%%%%%%%%%

 I reviewed these spacetimes representing ``uniformly accelerated objects'' in various places
 (see e.g. \cite{BISP,KVB} and references therein); 
 here I shall just mention some new results.
 
 The unique role of the boost-rotation symmetric spacetimes is exhibited by a
 theorem \cite{BPa} which roughly states that  in {axially} symmetric, locally
 asymptotically flat electrovacuum spacetimes (in the sense that a null infinity
 satisfying Penrose's requirements exists, but it need not
 necessarily exist globally), the only {additional} symmetry
 that does not exclude radiation is the {\it boost} symmetry.
 
 To prove such a result we start from the metric 
 
 \begin{eqnarray}
 \label{ds}
 ds^2&=&-{\left(r^{-1}\, V\, {{e}^{2\beta}}\!-\!r^2{e^{2\gamma }} U^2\,{\cosh 2\delta}
            \!-\!r^2{e^{-2\gamma}} W^2\,{\cosh 2\delta}\!-\!2r^2 UW\,{\sinh 2\delta}\right)}\ du^2\nonumber\\
     & &\ - 2{e^{2\beta}} du~dr
               -2r^2\left({e^{2\gamma }} U\,{\cosh 2\delta} +W\,{\sinh 2\delta}\right)\ du\ d\theta\nonumber\\
     & &    
         \ -2r^2\left({e^{-2\gamma}} W\,{\cosh 2\delta} +U\,{\sinh 2\delta}\right){\sin \theta}\ du\ d\phi\\
     & &\ + r^2\left[ {\cosh 2\delta}\left({e^{2\gamma }} { d}\theta^2 +{e^{-2\gamma}} {\sin^2 \theta}\ d\phi^2 \right)
                         +2\,{\sinh 2\delta}\ {\sin \theta}\ d\theta\ d\phi\right] \ ,\nonumber
 \end{eqnarray}
 where all funtions describing the metric and electromagnetic field tensor $F_{\mu \nu}$ are independent of $\phi$.
 Assuming asymptotic expansions of these functions at large $r$ with $u$, $\theta$, $\phi$ fixed 
 to guarantee asymptotic flatness, and using the outgoing radiation condition and the field equations,
 one finds the expansions to have specific forms. For example,
 
 $$
 \gamma  =\frac{c}{r}+( C-{{\scriptstyle{\frac{1}{6}}}}c^3
              -{{{\scriptstyle{\frac{3}{2}}}}} cd^2)
            \frac{1}{r^3}+...\ ,\ 
 ~~~~~~V   = r - 2M  + ...~,
 $$
 \begin{equation}
 ~~~~~~~~~F_{02} = X\;+\;(\epsilon_{,\theta}-e_{,u}){1\over r} + ...\ ,~~~~~~~~~~~~~F_{03} = Y\;-\;{f_{,u}\over r} + ...~,
 \label{jednarovnice}
 \end{equation}
 where the `coefficients' $c$, $d$, ... are functions of $u$ and $\theta$. The expansions are needed
 to further orders -- see \cite{BPa}     for their rather lengthy forms. Let us only recall that the decrease
 of the Bondi mass, 
 $
 m(u)=
 {1 \over 2}\int_{0}^{\pi} M(u,\theta)\sin \theta d\theta\ ,\label{hmota}
 $
 is given by
 \begin{equation}
 m_{,u} =-{{{\scriptstyle{\frac{1}{2}}}}}\int\limits_{0}^{\pi} (c,_u^2+d,_u^2+X^2+Y^2)\sin \theta d\theta
                \leq 0\ ,  \label{klhmota}
 \end{equation}
 where $c_{,u}$, $d_{,u}$, $X$, $Y$ are the gravitational and electromagnetic news functions.
 
 Now one writes down the Killing equations and solves them asymptotically in $r^{-1}$. One arrives 
 at the following theorem \cite{BPa}: Suppose that an axially symmetric
 electrovacuum spacetime admits a ``piece'' of ${\cal J}^+$ in the~sense
 that the~Bondi-Sachs coordinates can be introduced in which
 the~metric takes the~form (\ref{ds}), with
 the~asymptotic form of the~metric and electromagnetic field given by
 (\ref{jednarovnice}). If this spacetime admits an additional Killing
 vector forming with the~axial Killing vector a two-dimensional Lie
 algebra, then the additional Killing vector has asymptotically
 the~form 
 \begin{equation}
 \eta^\alpha=[-ku\cos \theta+\alpha(\theta),\
           kr\cos \theta+{\cal O}(r^{0}),\ -k\sin \theta+{\cal O}(r^{-1}),\ {\cal O}(r^{-1})]\ , 
 \label{Bbotr}
 \end{equation}
 where $k$ is a constant. For $k=0$ it generates asymptotically translations 
 (function $\alpha$ has then a specific form). For $k\not= 0$ it is the~boost Killing field.

 The case of translations is analyzed in detail in \cite{BAP}.
 Theorem 1, precisely formulated and proved there, states that if an
 asymptotically translational Killing vector is spacelike, then
 null infinity is singular at some $\theta \not= 0, \pi$; if it is
 null, null infinity is singular at $\theta = 0$ or $\pi$. The
 first case corresponds to cylindrical waves, the second case to a
 plane wave propagating along the symmetry axis. We refer to
 \cite{BAP} for the case when there is also a cosmic string
 present along the symmetry axis. The case of timelike Killing
 vector is described by Theorem 2 (proved also in \cite{BAP}):
 If an axisymmetric electrovacuum spacetime with a non-vanishing
 Bondi mass admits an asymptotically translational Killing vector
 and a complete cross section of ${\cal J}^+$, then the translational
 Killing vector is timelike and spacetime is thus stationary.
 
 The case of the boost Killing vector $(k\not=0)$ is thoroughly
 analyzed in \cite{BPa}. The general functional forms of the news
 functions (both gravitational and electromagnetic), and of the
 mass aspect and total Bondi mass of boost-rotation symmetric
 spacetimes are there given. Very recently these results were obtained \cite{VKR} by using the
 Newman-Penrose formalism and under more general assumptions 
 (for example, {$\cal J$} could in principle be polyhomogeneous).
 
 The general structure of the boost-rotation symmetric spacetimes
 with hypersurface orthogonal Killing vectors was analyzed in
 detail in \cite{ao}. Their radiative properties, including
 explicit  construction of radiation patterns and of Bondi mass
 for the specific boost-rotation symmetric solutions were
 investigated in several works --
 we refer to the reviews \cite{BISP,KVB} and \cite{PPRD}
 for details. There also the role of the boost-rotation symmetric
 spacetimes in such diverse fields like numerical relativity and
 quantum production of black-hole pairs is noticed and 
 references are given.
 
 Here I would like to mention yet a recent progress in
 understanding specific boost-rotation symmetric spacetimes with
 Killing vectors which are {\it not} hypersurface
 orthogonal. This is the {\it spinning} $C$-metric (see e.g. \cite{KSH}). It was
 discovered by Pleba\'nski and Demai\'nski as a generalization of the
 standard $C$-metric which is known to represent uniformly
 accelerated non-rotating black holes. In \cite{BPD} we first
 transformed the metric into Weyl coordinates, and then found 
 a transformation which brings it into the canonical form
 of the radiative spacetimes with the boost-rotation symmetry:
 
 \begin{eqnarray}
 \label{BStvarR}
 { ds}^2 &=& { e}^{\lambda} { d} \rho^2 + \rho^2 { e}^{-\mu} { d} \phi^2 \nonumber  \\
  & + & {(z^2-t^2)^{-1}} \left[ ({ e}^{\lambda} z^2 - { e}^{\mu} t^2 ) { d} z^2 \nonumber  
  -    2zt ({ e}^{\lambda} - { e}^{\mu}  ) { d} z\  { d} t 
 +     ({ e}^{\lambda} t^2 - { e}^{\mu} z^2 ) { d} t^2  \right]       \\ 
 ~ & - &    2{\cal A} { e}^{\mu} (z { d}t -  t { d} z)  { d} \phi -{\cal A}^2 { e}^{\mu} (z^2-t^2)   { d} \phi^2 \ ,
 \end{eqnarray}
 where functions $e^{\mu}, e^{\lambda}$ and ${\cal A}$ are given in terms of $(t,
 \rho, z)$ in a somewhat complicated but explicit manner. 
 This metric can represent two uniformly
 accelerated, spinning black holes, either connected by a conical
 singularity, or with conical singularities extending from each of
 them to infinity. The behaviour of the curvature
 invariants clearly indicates the presence of a non-vanishing
 radiation field (see Figure 5 in \cite{BPD}). The spinning
 $C$-metric is the only explicitly known example 
 with two Killing vectors which are not hypersurface orthogonal,
 in which one can give arbitrarily strong initial data on a hyperboloid ``above
 the roof" ($t>|z|$) which evolve into the radiative spacetime
 with smooth ${\cal J}^+$.

 \section{Inhomogeneous cosmologies and gravitational waves}
 
 Among the known vacuum inhomogeneous models, the {\it Gowdy solutions} (see e.g. \cite{KSH}) 
 have played the most distinct role. They belong to the class of solutions with 
 two commuting spacelike Killing vectors. Within a cosmological context, they form 
 a subclass of a wider class of $G_2$ {\it cosmologies} -- as are now commonly denoted  
 models which admit an Abelian group $G_2$ of isometries with orbits being spacelike 
 2-surfaces. A 2-surface with a 2-parameter isometry group must be a space of 
 constant curvature, and since neither a 2-sphere nor a 2-hyperboloid possess 
 2-parameter subgroups, it must be intrinsically flat. 
 If the 2-surface is an Euclidean plane or 
 a cylinder, then one speaks about planar or cylindrical universes. Gowdy universes 
 are compact -- the group orbits are 2-tori $T^2$.
 
 The metrics with two spacelike Killing vectors are often called the 
 generalized Einstein-Rosen metrics as, for example, by Carmeli, Charach and 
 Malin \cite{CCM} in their comprehensive survey of inhomogeneous cosmological 
 models of this type. In dimensionless coordinates ($t,z,x^1,x^2$), the line element can be 
 written as ($A,B = 1,2$) 
 \begin{equation}
 \label{Equx102}
 ds^2 / L^2 ~=~ e^F (-dt^2 + dz^2) + \gamma_{AB} dx^A dx^B,
 \end{equation}
 where $L$ is a constant length, $F$ and $\gamma_{AB}$  depend on $t$ and $z$ 
 only, and thus the spacelike Killing vectors are
 $
 {}^{(1)}\xi^\alpha= (0,0,1,0), {}^{(2)}\xi^\alpha= (0,0,0,1). 
 $
 
 Let us mention some recent developments in which the 
 Gowdy models have played a role. Gowdy-type models have been used to study the 
 propagation and collision of gravitational waves 
 with toroidal wavefronts in the FRW closed 
 universes with a stiff fluid \cite{BIJG}. In the standard Gowdy spacetimes it 
 is assumed that the ``twists'' associated with the isometry group on $T^2$ vanish. In 
 \cite{BCIM} the generalized Gowdy models without this assumption are considered, 
 and their global existence in time is proved. 
 
 As both interesting and non-trivial models, the Gowdy spacetimes have recently  
 attracted the attention of mathematical and numerical relativists.
 Chru\'{s}ciel, Isenberg and Moncrief \cite{GIM} proved that Gowdy spacetimes developed from a 
 dense subset in the initial data set cannot be extended past their singularities, 
 i.e. in ``most'' Gowdy models the strong cosmic censorship is satisfied.
 On cosmic censorship and spacetime singularities, especially in the 
 context of compact cosmologies, we refer to \cite{VMO}. This review shows 
 how intuition gained from such solutions as the Gowdy models or the Taub-NUT 
 spaces, when combined with new mathematical ideas and techniques, can produce 
 rigorous results with a generality out of reach until recently. To such 
 results belongs also the very recent work of Kichenassamy and Rendall \cite{KIR} 
 on the sufficiently general class of solutions (containing the maximum 
 number of arbitrary functions) representing unpolarized Gowdy spacetimes. The new 
 mathematical technique, the so called 
 Fuchsian algorithm, enables one to construct singular (and nonsingular) 
 solutions of partial differential equations with a large number of arbitrary 
 functions, and thus provide a description of singularities. Applying the Fuchsian 
 algorithm to Einstein's equations for Gowdy spacetimes with topology $T^3$, 
 Kichenassamy and Rendall have proved that general solutions behave at the (past) 
 singularity in a Kasner-like manner, i.e. they are asymptotically velocity 
 dominated with a diverging curvature invariant. One needs an 
 additional magnetic field not aligned with the two Killing vectors of the Gowdy 
 unpolarized spacetimes in order to get a general oscillatory (Mixmaster) 
 approach to a singularity, as shown by the numerical calculations \cite{WIB}. 
 
 Some metrics can be considered as exact {\it ``gravitational 
 solitons''} propagating on a cosmological background. 
 Verdaguer \cite{t} prepared a very complete review of solitonic cosmological
 solutions admitting two spacelike Killing vector fields.
 Recently, differential conservation laws for 
 large perturbations of gravitational field with respect to a given 
 curved background have been fomulated \cite{JOKB}. They should bring more light also on 
 various solitonic models in cosmology.
 
 %\vskip 3mm
 %I thank the organizers for inviting me to the interesting meeting and
 %Tom\'a\v{s} Ledvinka for the help with the manuscript.
 
 %%%%%%%%%%%%%%%%%%%%%%%%%%%%%%%%%%%%%%%%%%%%%%%%%%%%%%%%%%%%%%%%%%%%%%%%%% 
 %%%%%%%%%%%%%%%%%%%%%% Acknowledgements %%%%%%%%%%%%%%%%%%%%%%%%%%%%%%%%%%
 %%%%%%%%%%%%%%%%%%%%%%%%%%%%%%%%%%%%%%%%%%%%%%%%%%%%%%%%%%%%%%%%%%%%%%%%%%
 \vspace*{0.25cm} \baselineskip=10pt{\small \noindent 
 I thank the organizers for inviting me to the interesting meeting and
 Tom\'a\v{s} Ledvinka for the help with the manuscript. 
 Support from the grant No. GA\v CR 202/99/0261 of the Czech Republic is acknowledged.}


\begin{thebibliography}{288}
 \addcontentsline{toc}{section}{References}
 
 % 1. 
 \bibitem{BISP} Bi\v{c}\'{a}k, J. (2000) ``Selected solutions of Einstein's 
 field equations: their role in general relativity and astrophysics'',
 in ``Einstein's Field Equations and Their  Physical Implications'', 
 ed. B. G. Schmidt, Lecture Notes in Physics, Vol. 540, Springer Verlag
 
 % 2. 
 \bibitem{KVB} Bi\v{c}\'ak, J. (1997) in Relativistic Gravitation and Gravitational
 Radiation (Proceedings of the Les Houches School of Physics),
 eds. J.-A. Marck and J.-P. Lasota, Cambridge University Press,
 Cambridge
 
 % 3. 
 \bibitem{KSH} Kramer, D., Stephani, H., Herlt, E. and MacCallum,
 M. A. H. (1980) Exact solutions of Einstein's field equations,
 Cambridge University Press, Cambridge
 
 
 % 4. 
 \bibitem{JEK} Jordan, P., Ehlers, J. and Kundt, W. (1960) Akad. Wiss. Lit. Mainz, Abh. Math.
 Naturwiss. Kl., Nr. 2
 
 
 % 5. 
 \bibitem{AiB} Aichelburg, P. C., Balasin, H. (1996) Class. Quantum Grav. {\bf
 13}, 723
 
 
 % 6. 
 \bibitem{AiB2}  Aichelburg, P. C., Balasin, H. (1997) Class. Quantum
 Grav. {\bf 14}, A31
 
 
 % 7. 
 \bibitem{AS} Aichelburg, P. C., Sexl, R. U. (1971)  Gen. Rel. Grav. {\bf
 2}, 303
 
 
 % 8. 
 \bibitem{LoSa} Lousto, C. O., S\'anchez, N. (1989) Phys. Lett. {\bf B232}, 462
 
 
 % 9. 
 \bibitem{FePe} Ferrari, V., Pendenza, P. (1990)  Gen. Rel. Grav. {\bf 22}, 1105
 
 
 % 10. 
 \bibitem{BaNa} Balasin, H., Nachbagauer, H. (1995) Class. Quantum Grav. {\bf 12}, 707
 
 
 % 11. 
 \bibitem{PoGr1} Podolsk\'y, J., Griffiths, J. B. (1998) Phys. Rev. {\bf D58}, 124024
 
 
 % 12. 
 \bibitem{HoTa} Hotta, M., Tanaka, M. (1993) Class. Quantum
 Grav. {\bf 10}, 307
 
 
 % 13. 
 \bibitem{PoGr2} Podolsk\'y, J., Griffiths, J. B. (1997) Phys. Rev. {\bf D56}, 4756
 
 
 % 14. 
 \bibitem{PE} D'Eath, P. D. (1996) Black Holes: Gravitational
 Interactions, Clarendon Press, Oxford
 
 
 % 15. 
 \bibitem{Hoo} 't Hooft, G. (1987) Phys. Lett. {\bf B198}, 61
 
 
 % 16. 
 \bibitem{Fab} Fabbrichesi, M., Pettorino, R., Veneziano, G., 
 Vilkovisky, G. A. (1994) Nucl. Phys. {\bf B419}, 147
 
 
 % 17. 
 \bibitem{Kus1} Kunzinger, M., Steinbauer, R. (1999) Class. Quantum Grav. {\bf 16}, 1255
 
 
 % 18. 
 \bibitem{KuS} Kunzinger, M., Steinbauer, R. (1999) J. Math. Phys. {\bf 40}, 1479
 
 
 % 19. 
 \bibitem{PoVe} Podolsk\'y, J., Vesel\'y, K. (1998) Class. Quantum Grav. {\bf 15}, 3505
 
 
 % 20. 
 \bibitem{Per} Levin, O., Peres, A. (1994) Phys. Rev. {\bf D50}, 7421
 
 
 % 21. 
 \bibitem{Gi} Gibbons, G. W. (1999) Class. Quantum Grav. {\bf 16}, L 71
 
 
 % 22. 
 \bibitem{BiPr} Bi\v{c}\'{a}k, J., Pravda, V. (1998) Class. Quantum Grav. {\bf 15},
 1539
 
 
 % 23. 
 \bibitem{Gr} Griffiths, J. B. (1991) Colliding Plane Waves in General Relativity, Clarendon Press, Oxford
 
 
 % 24. 
 \bibitem{Yu1} Yurtsever, U. (1988) Phys. Rev. {\bf D38}, 1706
 
 
 % 25. 
 \bibitem{HET} Hauser, I., Ernst, F. J. (1989)
 J. Math. Phys. {\bf 30}, 872 and 2322;
 (1990) and (1991) J. Math. Phys. 
 {\bf 31}, 871 and {\bf 32}, 198
 
 
 % 26. 
 \bibitem{HET1} Hauser, I., Ernst, F. J. (1999) gr-qc/9903104
 
 
 % 27. 
 \bibitem{BIJG} Bi\v c\'ak, J.,  Griffiths, J. B. (1996)  Ann. Phys. (N.Y) {\bf 252}, 180
 
 
 % 28. 
 \bibitem{To} Tod, K. P. (1990) 
 Class. Quantum Grav. {\bf 7}, 2237
 
 
 % 29. 
 \bibitem{RI} d'Inverno, R. (1997) in Relativistic Gravitation and Gravitational Radiation,
 see \cite{KVB}
  
 
 
 % 30. 
 \bibitem{BER} Berger, B. K., Chru\'{s}ciel, P. T. and Moncrief,
 V. (1995) Ann. Phys. (N.Y.) {\bf 237}, 322
 
 
 % 31. 
 \bibitem{AP} Ashtekar, A., Pierri, M. (1996) J. Math.
 Phys. {\bf 37}, 6250
 
 
 % 32. 
 \bibitem{ABS1} Ashtekar, A., Bi\v{c}\'{a}k, J. and Schmidt, B. G.
 (1997) Phys. Rev. {\bf D55}, 669
 
 
 % 33. 
 \bibitem{ABS2}  Ashtekar, A., Bi\v{c}\'{a}k, J. and Schmidt, B. G.
 (1997) Phys. Rev. {\bf D55}, 687
 
 
 % 34. 
 \bibitem{ak}Chru\'{s}ciel, P. T. (1992)
 Proc. Roy. Soc. Lond. A {\bf 436}, 299;
 Chru\'{s}ciel, P. T., Singleton, D. B. (1992) 
 Commun. Math. Phys. {\bf 147}, 137, and references therein
 
 
 % 35. 
 \bibitem{BiPo} Bi\v{c}\'{a}k, J., Podolsk\'{y}, J. (1997) Phys. Rev. {\bf D55}, 1985
 
 
 
 % 36. 
 \bibitem{al} Bi\v{c}\'{a}k, J., Podolsk\'{y}, J. (1995) Phys. Rev. {\bf D52}, 887
 
 
 
 
 
 
 
 
 
 % 37. 
 \bibitem{STP} Stephani, H. (1993) Class. Quantum Grav. {\bf 10}, 2187
 
 % 38. 
 \bibitem{FIL} Finley, J. D., Pleba\'nski, J. F. and Przanowski, M. (1997)
 Class. Quant. Grav. {\bf 14}, 487 
 
 % 39. 
 \bibitem{MA} MacAlevey, P.  (1999) Class. Quantum Grav. {\bf 16}, 2259
 
 % 40. 
 \bibitem{MASO} Mason, L. (1998) Class. Quantum Grav. {\bf 15}, 1019
 
 % 41. 
 \bibitem{BPa} Bi\v{c}\'{a}k, J., Pravdov\'{a}, A. (1998)
 J. Math. Phys. {\bf 39}, 6011
 
 
 % 42. 
 \bibitem{BAP} Bi\v{c}\'{a}k, J., Pravdov\'{a}, A. (1999)
 Class. Quantum Grav. {\bf 16}, 2023
 
 
 % 43. 
 \bibitem{VKR} Valiente-Kroon, J. A. (2000)
 J. Math. Phys., {\bf 41}, 898
 
 % 44. 
 \bibitem{ao} Bi\v{c}\'{a}k, J., Schmidt, B. G. (1989)
 Phys. Rev.  {\bf D40}, 1827
 
 
 % 45. 
 \bibitem{PPRD} Pravda, V., Pravdov\'a, A. (2000) Czech. J. Physics {\bf 50}, 333
 
 % 46. 
 \bibitem{BPD} Bi\v{c}\'ak, J., Pravda, V. (1999) 
 Phys. Rev. {\bf D60}, 044004
 
 
 % 47. 
 \bibitem{CCM} Carmeli, M., Charach, Ch. and Malin, S. (1981) 
 Physics Reports {\bf 76}, 79
 
 
 % 48. 
 \bibitem{BCIM} Berger, B. K., Chru\'sciel, P. T., Isenberg, J., Moncrief, V. (1997) 
 Ann. Phys. (N.Y.) {\bf 260}, 117
 
 
 % 49. 
 \bibitem{GIM} Chru\'sciel, P. T., Isenberg, J. and Moncrief, V. (1990) 
 Class. Quantum Grav. {\bf 7}, 1671
 
 
 % 50. 
 \bibitem{VMO} Moncrief, V. (1997) 
 in Proc. of the 14th International Conference on General Relativity and Gravitation,
 eds. M. Francaviglia, G. Longhi, L. Lusanna and E. Sorace, World Scientific, Singapore
 
 
 % 51. 
 \bibitem{KIR} Kichenassamy, S., Rendall, A. D. (1998) Class. Quantum Grav. {\bf 15}, 1339
 
 
 % 52. 
 \bibitem{WIB} Weaver, M., Isenberg, J. and Berger, B. K. (1998) 
 Phys. Rev. Lett. {\bf 80}, 2984
 
 
 % 53. 
 \bibitem{t} Verdaguer, E. (1993) Physics Reports {\bf 229}, 1
 
 
 % 54. 
 \bibitem{JOKB} Katz, J., Bi\v c\'ak, J. and Lynden-Bell, D. (1997) 
 Phys. Rev. {\bf D55}, 5957
  
 
 \end{thebibliography}
 \end{document}